\documentclass{article}

\usepackage{PRIMEarxiv}
\usepackage[numbers]{natbib}
\usepackage{amsmath}
\usepackage[utf8]{inputenc} 
\usepackage[T1]{fontenc}    
\usepackage{hyperref}       
\usepackage{url}            
\usepackage{booktabs}       
\usepackage{amsfonts}       
\usepackage{nicefrac}       
\usepackage{microtype}      
\usepackage{lipsum}
\usepackage{fancyhdr}       
\usepackage{graphicx}       
\graphicspath{{media/}}     

\title{Know Your Scientist: KYC as Biosecurity Infrastructure}

\author{
  Jonathan Feldman \\
  Georgia Institute of Technology \\
  Atlanta, GA\\
  \texttt{jonathanfeldman@gatech.edu} \\
   \And
  Tal Feldman \\
  Yale Law School \\
  New Haven, CT\\
  \texttt{tal.feldman@yale.edu} \\
 \And
  Annie I. Anton \\
  Georgia Institute of Technology \\
  Atlanta, GA\\
  \texttt{aa16@gatech.edu} \\
}

\begin{document}
\maketitle

\begin{abstract}
Biological AI tools for protein design and structure prediction are advancing rapidly, creating dual-use risks that existing safeguards cannot adequately address. Current model-level restrictions, including keyword filtering, output screening, and content-based access denials, are fundamentally ill-suited to biology, where reliable function prediction remains beyond reach and novel threats evade detection by design. We propose a three-tier Know Your Customer (KYC) framework, inspired by anti-money laundering (AML) practices in the financial sector, that shifts governance from content inspection to user verification and monitoring. Tier I leverages research institutions as trust anchors to vouch for affiliated researchers and assume responsibility for vetting. Tier II applies output screening through sequence homology searches and functional annotation. Tier III monitors behavioral patterns to detect anomalies inconsistent with declared research purposes. This layered approach preserves access for legitimate researchers while raising the cost of misuse through institutional accountability and traceability. The framework can be implemented immediately using existing institutional infrastructure, requiring no new legislation or regulatory mandates.
\end{abstract}


\section{Introduction}

Advances in artificial intelligence are rapidly expanding the capabilities of biological research tools. Systems for structure prediction, sequence modeling, and generative protein design now support tasks that were recently limited to specialized laboratories \cite{ruffolo_designing_2024,CHEN2023706,xiao2025proteinlargelanguagemodels,NEURIPS2021_f51338d7,hie_evolutionary_2022,Huot2025.05.12.653592, af3complex}. These tools promise substantial benefits for medicine and biotechnology, but they also introduce dual-use risks that existing safeguards cannot address.

Most current approaches focus on model-level restrictions—keyword filtering, output screening, or access denials based on content \cite{doi:10.1126/science.adu8578, feldman_resilient_2025}. While intuitive, these methods are poorly suited to biology. Keyword filtering can be circumvented 
\cite{feldman_resilient_2025, marsoof_content-filtering_2023, ball2025impossibilityseparatingintelligencejudgment}. Sequence homology detects known pathogens but misses novel designs by construction \cite{doi:10.1126/science.adu8578}. Functional annotation remains incomplete, and reliable prediction of pathogenicity or toxicity is beyond current capabilities \cite{feldman_resilient_2025, dianzhuo_wang_without_2025, LIU2025100964}. As biological design tools grow more powerful, these limitations create a difficult tradeoff: restrict access so heavily that legitimate research is impeded, or permit access so broadly that misuse becomes difficult to detect or attribute.

But there is another way. Instead of asking solely whether a particular output is dangerous—a question current systems cannot reliably answer \cite{feldman_resilient_2025, 10.1242/dmm.052218}—one can ask whether the user generating that output is one we trust. This shift from content inspection to user verification reflects the technical limits of biological function prediction, but it does not come from us alone. Rather, it is inspired by the analogous regulatory system created in the financial sector. 

Indeed, Anti-Money Laundering (AML) frameworks from the financial sector provide a guide to how a similarly high-risk and complex problem was addressed, implementing the logic that we argue for here \cite{bcbs_consolidated_kyc_2004, fed_supervision_manual_kyc, fincen_cdd_final_rule_fedreg_2016, ffiec_bsaaml_manual, finra_rule_2090_kyc, deepak_review_nodate, chen_you_2020}. AML systems aim to prevent the illicit flow of funds-for terrorism, organized crime, sanctions evasion, and the like-through layered identity verification of the users and behavioral monitoring of their actions \cite{bcbs_consolidated_kyc_2004, ffiec_bsaaml_manual, fatf_recommendations_2025}. At the root of this regime are the Know Your Customer (KYC) requirements. They require that financial institutions verify customer identity and refuse service to sanctioned entities before providing access \cite{egan_oversight_2023, fincen_cdd_final_rule_fedreg_2016, turn0search2}. Transaction monitoring further detects suspicious patterns over time \cite{ibm_aml_transaction_monitoring, ffiec_bsaaml_manual, bcbs_consolidated_kyc_2004}. The risk AML aims to prevent—enabling financial crime—is instructive for addressing the risk in biological AI: enabling pathogen design.

Although KYC-style approaches have been mentioned in AI governance discussions, they have not been specified in an operational, biology-specific manner \cite{egan_oversight_2023, tarangelo_protecting_2025, facini_biosecurity_2025, moulange2023responsiblegovernancebiologicaldesign}. Existing proposals remain at the level of single-sentence recommendations rather than implementable architectures. This paper fills that gap.

The timing is urgent. Protein design capabilities have advanced rapidly over the past three years, with systems like AlphaFold, ESM3, and RFdiffusion enabling tasks that were recently out of reach \cite{ruffolo_designing_2024,CHEN2023706,xiao2025proteinlargelanguagemodels,watson_novo_2023, huss_mapping_2021, King2025.09.12.675911}. Many of these tools are now accessible through APIs, creating natural chokepoints for access control \cite{esm_faqs, af3server_alphafold, feldman_position_nodate}. Scientists have called for stronger safeguards \cite{wang_call_2025, doi:10.1126/science.ado1671}, yet concrete, implementable frameworks remain scarce.

\begin{figure}[h]
\centering
\includegraphics[width=0.7\textwidth]{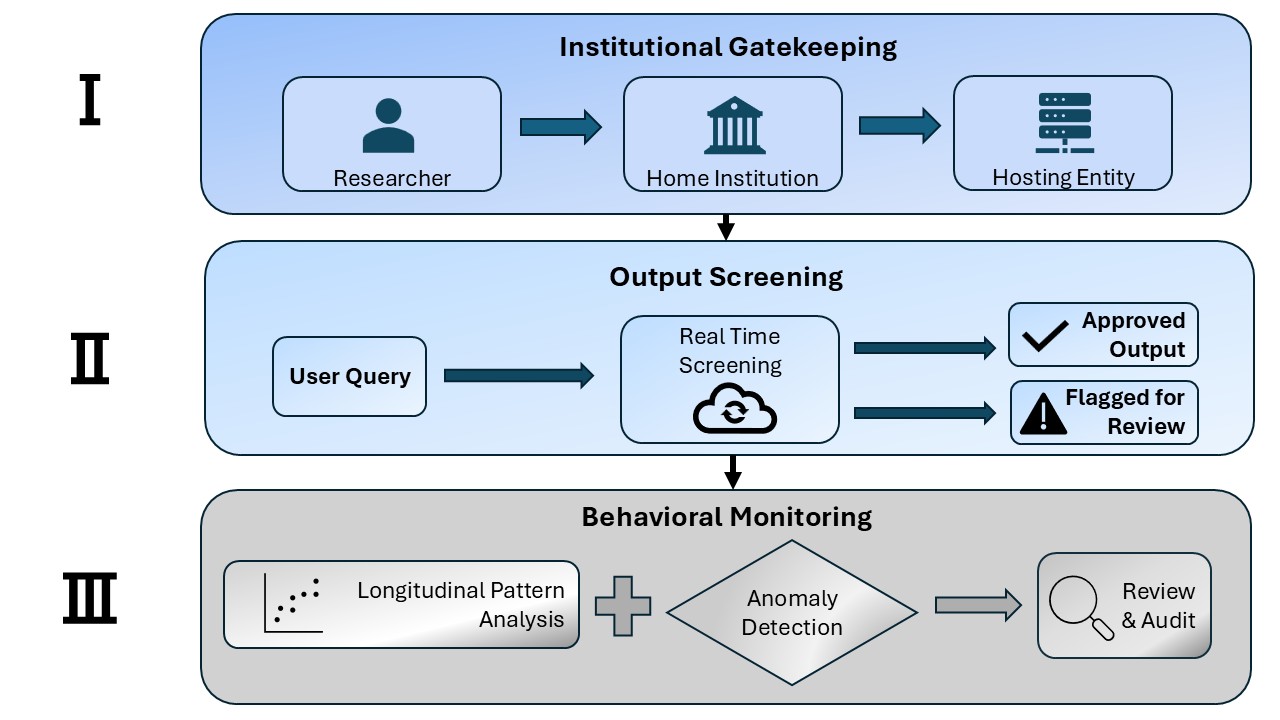}
\caption{An architectural illustration of the three-tier KYC framework.  \textbf{Tier I (Institutional Gatekeeping):} Research institutions vouch for affiliated users and assume accountability for vetting. \textbf{Tier II (Output Screening):} Real-time analysis of generated sequences using homology searches and functional annotation. \textbf{Tier III (Behavioral Monitoring):} Longitudinal pattern analysis detects activity inconsistent with declared research purposes. Each tier provides independent security value while preserving access for legitimate researchers.}
\label{fig:kyc_architecture}
\end{figure}

Hence, we propose a three-tier KYC framework (Figure~\ref{fig:kyc_architecture}) that leverages existing research institutions as trusted anchors, augments access control with output and behavioral monitoring, and establishes meaningful consequences for misuse or negligent oversight. This approach preserves access for legitimate researchers, raises the cost of misuse, and can be implemented immediately using existing institutional infrastructure—without waiting for new regulatory mandates.

\section{Three-Tier KYC Framework}

The framework adapts the layered structure of financial AML compliance to biological AI. In AML systems, KYC establishes verified identity at entry, while ongoing monitoring detects suspicious behavioral patterns \cite{bcbs_consolidated_kyc_2004, ffiec_bsaaml_manual, ibm_aml_transaction_monitoring}. Neither mechanism alone provides adequate protection—identity verification without monitoring cannot detect gradual misuse; monitoring without verified identities cannot enable attribution. We implement the same layered approach: Tier I establishes verified researcher identity and institutional accountability. Tiers II-III monitor ongoing activity through output screening and behavioral pattern analysis.

The design leverages existing institutional infrastructure rather than introducing new layers of bureaucracy. Research institutions already vet researchers through hiring, training, and oversight processes, including biosafety and ethics review \cite{gao_biosafety_2025, grady_institutional_2015, serpico_institutional_2022, noauthor_institutional_2025}. These mechanisms provide a natural basis for access control, allowing model hosts to rely on institutional judgment rather than attempting to infer legitimacy or intent from model interactions alone.

Minimizing friction for legitimate users is a core constraint. Most biological design tools require downstream wet-lab validation to have any practical impact, which in turn requires institutional affiliation or formal collaboration  \cite{feldman_resilient_2025, dianzhuo_wang_without_2025, suveena_translational_2025, King2025.09.12.675911, Stark2025.11.20.689494}. Aligning access with these existing structures preserves usability for legitimate research while raising the barrier to misuse.

Finally, the framework emphasizes accountability through consequences rather than brittle algorithmic restrictions. Automated filters with high false-positive rates are poorly suited to governing biological design. In contrast, clear attribution, traceability, and enforceable penalties create durable incentives for responsible use. These principles motivate a layered architecture in which each tier contributes independent security value while reinforcing the others.

\subsection{Tier I: Institutional Gatekeeping}

The first tier establishes trust at the point of access. Researchers seeking to use a biological AI tool apply through their home institution rather than directly to the model provider. The institution evaluates the request, confirms the researcher's identity and role, and vouches for the legitimacy of the proposed use. The hosting entity, in turn, maintains a list of trusted institutions whose endorsements it accepts and grants access based on institutional approval.

Throughout this framework, research institutions refer to organizations that conduct biological research under recognized oversight structures. This includes universities and colleges, government research laboratories, non-profit research institutes, hospital-affiliated research centers, and corporate R\&D divisions. The common thread is institutional accountability: many of these organizations maintain biosafety committees, ethics review boards, or similar governance structures, and all employ researchers under formal relationships. These oversight mechanisms provide a foundation that can be leveraged for managing access to biological AI tools.

This is similar to KYC's role in AML compliance. Just as banks verify customer identity before providing financial services and maintain exclusion lists identifying prohibited actors \cite{bcbs_consolidated_kyc_2004, fincen_cdd_final_rule_fedreg_2016, ffiec_bsaaml_manual} biological AI hosts verify researcher identity through institutional vouching. The institution functions as the compliance officer: responsible for confirming the researcher is who they claim, their intended use is legitimate, and they do not appear on government-maintained exclusion lists \cite{ofac_sdn_list, state_dept_fto_list}.

This arrangement shifts the burden of vetting from model providers to institutions that are better positioned to assess researcher qualifications and intentions. It also creates a clear chain of accountability. Institutions that vouch for researchers implicitly accept responsibility for their vetting, and misuse by affiliated users carries reputational and access consequences.

AML frameworks rely on objective exclusion criteria rather than subjective trustworthiness judgments. Regulators maintain lists of sanctioned individuals, designated terrorist entities, and convicted financial criminals \cite{ofac_sdn_list, state_dept_fto_list, un_sc_consolidated_list}. Biological AI governance adopts the same approach. Hosting entities should deny access to individuals or organizations appearing on government-maintained exclusion lists, including OFAC’s Specially Designated Nationals, the State Department’s Foreign Terrorist Organizations list, and the Commerce Department’s Entity List, as well as individuals convicted under biological weapons statutes\cite{ofac_sdn_list, state_dept_fto_list, doc_entity_list}. Beyond these bright-line exclusions, institutions retain discretion for context-specific vetting—analogous to enhanced due diligence banks conduct for higher-risk customers \cite{fatf_recommendations_2025, bcbs_consolidated_kyc_2004}.

\subsubsection{Outlining Inclusion Criteria}

Criteria for institutional trust are determined by the hosting entity and scale with the assessed risk of the model or tool to which access is being granted. Each host evaluates prospective institutions according to its own trust framework, which may include factors such as the presence of an Institutional Biosafety Committee (IBC) \cite{noauthor_institutional_2025, egan_oversight_2023}, receipt of federally funded research subject to established oversight regimes, prior compliance history, or an existing relationship with the host. Crucially, these institutional signals are interpreted in light of the model’s risk profile, rather than in isolation.

Ongoing work on biological AI risk assessment increasingly enables models to be classified by their potential for misuse or harm, allowing hosts to calibrate trust requirements accordingly \cite{de_haro_biosecurity_2024, ackerman_biothreat_2025, noauthor_global_nodate}. For lower-risk systems—such as small protein language models comparable to those in the ESM family \cite{hayes_simulating_2024}—the threshold for institutional trust may reasonably be lower. In contrast, access to more advanced systems, such as models capable of de novo genome design \cite{hwang_genomic_2024, brixi_genome_2025}, may warrant substantially stricter requirements, effectively limiting access to institutions with demonstrated oversight capacity and clearly legitimate use cases.

To accommodate common research practices within this risk-sensitive framework, institutions may sponsor non-affiliated individuals, including visiting scholars, collaborators, or independent researchers. In such cases, the sponsoring institution submits the access request, vets both the individual and the proposed use, and assumes responsibility for ongoing oversight. Any misuse would affect the sponsor’s standing with hosting entities, reinforcing accountability while preserving flexibility. Importantly, the willingness—or refusal—of institutions to sponsor a given individual provides an additional signal of perceived risk, further integrating model risk, institutional trust, and host-specific requirements into a unified access control framework.

\subsection{Tier II: Output Screening}

The second tier introduces continuous monitoring of model outputs to detect potentially concerning activity. Generated sequences or designs are analyzed in real time using available screening tools. Current methods include sequence homology searches, such as BLAST, and ontology-based functional annotation to identify similarity to known pathogens or hazardous functions \cite{doi:10.1126/science.adu8578, feldman_resilient_2025, Zhao2025.05.07.652715, miller_post-processing_1997}. These tools are imperfect, but they provide useful signals when interpreted in context.

As predictive technologies advance, this tier can incorporate more sophisticated tools, including protein function predictors and toxicity assessments \cite{feldman_resilient_2025, feldman_position_nodate, wang_deep-learning-enabled_2023}. The framework is intentionally modular, allowing screening capabilities to evolve without altering the underlying access-control structure. In addition, emerging watermarking and output-provenance technologies for generative models can be integrated into these pipelines, providing an extra layer of accountability—particularly when generated sequences are synthesized or otherwise deployed \cite{chen_enhancing_nodate, zhang_foldmark_2024}.

When outputs are flagged as potentially pathogenic, viral, or otherwise concerning, and the user's declared use case does not justify access to such outputs, an inquiry can be initiated. Flagged outputs may be temporarily frozen and held for review, and the associated user activity is logged. Crucially, the focus is not on automated enforcement but on generating signals that prompt human oversight. Users are aware that their activity is monitored and attributable, reinforcing accountability \cite{noauthor_user_nodate}.

For example, a researcher developing vaccine candidates against influenza may legitimately generate sequences resembling viral proteins. The system would flag these outputs due to their similarity to pathogen components, but a review would confirm the researcher’s declared use case of vaccine development, allowing access to continue. In contrast, if a user whose stated purpose is analyzing mitochondrial genomes produces similar sequences, the discrepancy between their actions and declared use case would trigger an escalated review.

This mirrors AML transaction monitoring, which generates alerts and escalates potentially suspicious activity for review rather than relying on fully automated enforcement. Monitoring systems are designed to identify patterns or anomalies that warrant investigation and contextual assessment, and flagged activity may proceed if review determines it to be consistent with legitimate purpose \cite{ffiec_bsaaml_manual, fatf_recommendations_2025}. The objective is not perfect automated detection—which AML frameworks explicitly recognize as unattainable—but the generation of signals that enable human oversight, accountability, and attribution \cite{bcbs_consolidated_kyc_2004, ffiec_bsaaml_manual}. Similarly, pathogen-homologous outputs may trigger review without necessarily resulting in access denial when a legitimate research use, such as vaccine development, is documented \cite{feldman_resilient_2025}.

\subsection{Tier III: Behavioral Pattern Monitoring}

The third tier addresses risks that emerge over time rather than from individual outputs. User activity is analyzed longitudinally to identify patterns that deviate from declared use cases or expected research behavior. Signals may include repeated interactions with flagged outputs, aggregate risk signals over time, unexpected access patterns, or other inconsistencies between stated research goals and observed usage. This approach is analogous to AML behavioral monitoring \cite{nist_sp800_63b_4, ffiec_authentication_access_2021, ibm_aml_transaction_monitoring, ffiec_bsaaml_manual}.

Rather than relying on single events, this tier uses threshold-based accumulation of signals to trigger review. Importantly, decisions to suspend or revoke access are not automated. Human review should be required before enforcement actions are taken, allowing contextual judgment and reducing the risk of erroneous penalties.

Together, these three tiers form a deep defense framework. Institutional gatekeeping establishes trust at entry, output screening provides immediate response, and behavioral monitoring captures longer-term risk. Each tier compensates for the limitations of the others, creating a system that is both practical and robust as biological AI capabilities continue to advance.

\subsection{Shared Responsibility as the Foundation of the Framework}

Effective governance of biological AI systems requires a genuinely shared allocation of responsibility between research institutions and hosting entities. Research institutions serve as the first line of defense by vetting applicants, evaluating proposed use cases, and ensuring that researchers operate within appropriate oversight structures. This institutional gatekeeping is essential, but it is inherently limited by the institution’s visibility into downstream model use and therefore cannot function as a complete safeguard on its own.

Hosting entities bear an independent and complementary responsibility that research institutions cannot fulfill alone. Hosts must assess which institutions they trust, calibrate access requirements to the risk profile of the model, and implement technical oversight mechanisms such as output screening, activity logging, and behavioral monitoring. Because hosts have direct, system-level visibility into model interactions, they are uniquely positioned to detect anomalous usage patterns or emergent risks that may not be observable at the institutional level.

The framework is effective precisely because responsibility is distributed rather than centralized. Neither research institutions nor hosts are individually capable of managing the full risk surface of advanced biological AI systems. Instead, institutions evaluate researchers and intent, while hosts validate institutions and monitor interaction with the technology. This shared and interdependent oversight structure minimizes failure, reinforces accountability on both sides, and ensures that risk management remains an ongoing, collective obligation.

\section{Adoption Incentives}

\subsection{Researcher Acceptance}

Preventing biosecurity risk currently requires some access friction—it is impossible to deter misuse without restrictions \cite{anderljung2023protectingsocietyaimisuse}. The question is whether friction remains tolerable for legitimate users. Several factors support researcher acceptance of the framework proposed here.

First, friction is minimal. Researchers already navigate institutional approvals for laboratory work, grants, and protocols \cite{gao_biosafety_2025, HOFFMANN2023106165, tang_enhancing_2024}. A one-time institutional access request represents marginal additional burden. Most researchers continue working without interruption after initial approval.

Second, restrictions target those who should not have unrestricted access. Unlike popular LLMs which are general-purpose and can be used by a wide swath of the population for legitimate uses, there is little reason for individuals who are not doing cutting-edge biological research and have no institutional affiliation to independently access tools capable of designing pandemic-potential agents \cite{wang_call_2025, dianzhuo_wang_without_2025, feldman_opinion_2025}—just as they should not work unsupervised with live biological agents \cite{noauthor_institutional_2025, gao_biosafety_2025}.

Third, the AML analogy is instructive. Banks verify every customer's identity and subject customer transactions to ongoing monitoring, introducing friction for account holders worldwide \cite{fincen_cdd_final_rule_fedreg_2016, ffiec_bsaaml_manual, bcbs_consolidated_kyc_2004}. Despite these requirements, modern payment systems operate at global scale, processing large volumes of transactions each day \cite{fed_payments_study_frps, fed_payments_study_npips_2021_22}. Risk-based AML and KYC controls are designed to be proportionate so they do not unnecessarily disrupt legitimate activity, while monitoring and suspicious activity reporting increase the detection, traceability, and disruption of illicit finance \cite{fatf_risk_based_approach}. 

\subsection{Hosting Entity Incentives}

For hosting entities, this framework is far more tenable than the speculative alternative: building systems that reliably predict whether designed proteins are dangerous \cite{feldman_resilient_2025, dianzhuo_wang_without_2025}. Current biological AI cannot determine sequence pathogenicity reliably \cite{feldman_resilient_2025, cao_prediction_2025}. Solving this requires breakthroughs in structural biology, immunology, and machine learning—an expensive, uncertain, potentially impossible task \cite{doi:10.1126/science.adu8578, de_haro_biosecurity_2024, gao_biosafety_2025}.

KYC resolves this by deferring vetting and attribution to institutions, reducing the risk exposure for hosting entities. The institution vouches for a user, the user operates under institutional oversight, and misuse traces to both. The hosting entity verifies institutional credentials and maintains access records.

Reputational incentives also favor adoption. High-profile misuse of a host's tool could trigger regulatory backlash and institutional distrust. Proactive safeguards function as reputational insurance \cite{BRAVOURQUIZA2021101344, chukwudi_how_2025}.

\section{Enforcement Mechanisms and Regulatory Pathways}

\subsection{Voluntary Adoption with Distributed Enforcement}

The framework is designed for immediate voluntary adoption by hosting entities without requiring regulatory mandates. Companies providing API access to biological design tools—including AlphaFold Server, ESM APIs, RFdiffusion implementations, and similar platforms—can implement institutional gatekeeping, output screening, and behavioral monitoring unilaterally through terms of service \cite{esm_faqs, af3server_alphafold, zhang_generative_2025, 10.1145/3375627.3375873}. This voluntary approach enables rapid deployment while regulatory pathways develop.

The primary enforcement mechanism is access revocation. Biological AI tools represent powerful capabilities that accelerate research and enable designs previously requiring specialized laboratory infrastructure \cite{ruffolo_designing_2024, watson_novo_2023, zhang_generative_2025}. Access to these tools is a privilege, not a right. Users who violate terms of service through misrepresentation, circumvention of safety measures, or generation of outputs inconsistent with declared research purposes lose access. Institutions that repeatedly vouch for problematic users or fail to address misconduct by affiliated researchers risk institution-wide access revocation.

Critically, enforcement gains strength through information sharing between hosting platforms. Just as financial institutions share information about suspicious actors through regulatory reporting systems and authorized inter-institution information-sharing mechanisms \cite{ffiec_sar_overview, fincen_314b, noauthor_how_nodate}, biological AI hosts can establish shared exclusion lists. When one platform revokes access due to misuse, that individual or institution appears on a shared registry accessible to other platforms. 

This distributed enforcement structure addresses a key challenge: individual platforms may hesitate to revoke access due to competitive pressure or uncertainty about violations. If problematic users can immediately migrate to alternative platforms, enforcement loses deterrent value. Shared exclusion lists eliminate this escape route, making violations consequential. An institution that loses trusted status with major platforms effectively loses access to the biological AI ecosystem, creating strong incentives for rigorous internal oversight.

\subsection{Existing Models of Delegated Oversight}

U.S. biosecurity and research security governance has often relied on institutional intermediaries rather than comprehensive, individualized federal vetting of all researchers. In several high-risk domains, the federal government establishes baseline rules while conditioning participation in the regulated activity on institutions assuming responsibility for day-to-day access control and oversight \cite{ecfr_42cfr73_select_agents_2025} \cite{fsap_security_risk_assessment_2024}. This division of labor reflects practical constraints: federal agencies set guardrails and enforce compliance, but institutions translate those requirements into operational decisions about who may access sensitive capabilities and under what conditions.

The Federal Select Agent Program exemplifies this hybrid model \cite{fsap_annual_report_2024, select_agent_regulations}. Federal law determines which biological agents are regulated and identifies individuals who are legally barred from access \cite{select_agent_regulations, fsap_annual_report_2024}. At the same time, research entities must register, designate responsible officials, and maintain internal controls governing access to regulated materials \cite{select_agent_regulations, fsap_annual_report_2024}. Individual access decisions occur within the institution, informed by—but not supplanted by—federal security risk assessments conducted through the Department of Justice \cite{fsap_security_risk_assessment_2024, fsap_annual_report_2024, select_agent_regulations}. The government thus retains authority over exclusion criteria and enforcement, while institutions bear primary responsibility for vetting, monitoring, and controlling access in practice \cite{select_agent_regulations, fsap_annual_report_2024, fsap_inspection_summary_2024}.

A similar structure appears in export control and research security regimes. This typically involves screening potential partners and visitors against government-maintained restricted lists, assessing whether planned activities implicate controlled items or technology, and establishing internal procedures or safeguards to manage compliance risks \cite{mit_restricted_party, rochester_export_compliance, bu_export_control_manual}. Federal agencies define what is controlled and when a license is required, but institutions handle day-to-day compliance and can be penalized for violations. Across export controls and research security, this split—federal rules and enforcement paired with institutional gatekeeping—has become a common way to manage sensitive research in complex settings.

\subsection{Federal Pathways for Standardization}

While voluntary adoption enables immediate implementation, federal action could accelerate adoption and establish consistent standards across providers. Several pathways exist under current executive authority.

Federal research funding agencies already impose biosafety and dual-use research oversight requirements on recipient institutions \cite{usg_durc_pepp_policy_2024, nih_durc_policy, gao_biosafety_2025}. Agencies could extend these requirements to include verification that researchers accessing biological AI tools operate under institutional oversight frameworks meeting minimum standards. The NIH, NSF, and DOE, who are major funders of U.S. biological research, could condition awards on institutional implementation of KYC-style access controls for affiliated researchers \cite{noauthor_federal_nodate, noauthor_fact_2022}. This would likely not require new legislation and could be achieved through administrative rulemaking specifying that institutions receiving federal funds must establish processes for vetting researcher access to biological design tools and maintaining accountability for their use.

The National Institute of Standards and Technology (NIST), through CAISI, maintains authority to develop voluntary consensus standards for AI systems \cite{noauthor_frameworks_2017, noauthor_standards_2016}. NIST could convene stakeholders—hosting entities, research institutions, biosecurity experts—to develop technical standards for implementing the three-tier framework. These standards could be voluntary but would likely become de facto industry requirements through market pressure and federal procurement specifications. They may also become binding rules through agency rulemaking. 

Notably, the federal government has already moved to make biosecurity a priority, focusing on nucleic acid synthesis screening. The Biden Administration's 2024 Screening Framework Guidance for Providers of Synthetic Nucleic Acids establishes requirements for DNA synthesis companies to screen orders against databases of concerning sequences and verify customer identity \cite{us_nucleic_acid_synthesis_framework_2024, hhs_synthetic_nucleic_acid_guidance_2023}. This framework recognizes that biological capability risks require verification of who has access, not just what sequences are synthesized. Our proposed KYC framework represents the logical upstream extension: if we screen synthesized DNA orders, we should also govern access to the AI tools that design those sequences in the first place.

The Trump administration's AI policy discussions have similarly emphasized synthesis screening as a national security priority, with proposals to expand screening requirements and enhance enforcement mechanisms \cite{americas_ai_action_plan_2025}. Extending governance upstream to biological AI design tools aligns with this trajectory. AI-generated designs must eventually be synthesized to pose physical risks \cite{feldman_resilient_2025, MOUNEYRAC2017241, suveena_translational_2025}, but governing only at the synthesis stage creates gaps—researchers can accumulate dangerous knowledge, share designs through alternative channels, or synthesize abroad. Comprehensive biosecurity requires governance at both the design and synthesis stages \cite{feldman_resilient_2025,doi:10.1126/science.adu8578, doi:10.1126/science.ado1671}.

Federal action could also address the open-weight model problem. While this framework cannot control locally-run models, agencies could condition federal funding or establish export controls for the most capable biological design models, ensuring they remain under hosted access with implemented safeguards. This would not prevent all capability diffusion but would slow proliferation of the highest-risk tools \cite{feldman_position_nodate}.

\subsection{Federal Oversight for High-Risk Biological AI}

As biological AI tools increase in capability, the risks they pose may reach a level that necessitates direct federal oversight. Implementing this pathway is likely the most challenging option, requiring dedicated funding, infrastructure, and sustained administrative capacity. In such cases, the federal government could assume a role in vetting individual researchers, either independently or in coordination with research institutions and hosting entities. This responsibility could be managed by an existing agency or a dedicated subsidiary, such as within the Department of Energy or the Centers for Disease Control.

Under this model, hosting institutions would continue to monitor usage and track user activity, while the federal entity would oversee researcher vetting. The government could maintain a registry of high-risk models and establish statutory, funding-related, or collaborative obligations—such as for federal contractors—to ensure compliance with vetting procedures.

Recognizing practical limits on resources, federal vetting could focus on researchers seeking access to the riskiest models, as defined by formal risk assessments. Importantly, vetting would remain a shared responsibility: research institutions could continue to serve as primary evaluators, with the federal government providing secondary review and oversight. While this approach is more difficult to implement than host-level or collaborative oversight, it provides the most rigorous multi-layered safeguards for the highest-risk models, enabling scalable and accountable governance.

\section{Limitations and Open Questions}

The framework has several important limitations. Open-source models bypass access control entirely, rendering KYC ineffective without centralized hosting \cite{feldman_position_nodate, black2025openweightgenomelanguagemodel}. To apply this framework, biological AI tools must not be made open source. 

KYC is also not a complete defense against determined, well-resourced state actors that can establish front institutions. Its primary strength lies in deterring lone actors and raising the cost of misuse through accountability and traceability, rather than eliminating all possible threats. Still, the instructive AML example from finance shows that there are ways to detect even well-resourced actors that hide behind complex ownership structures. 

Calibration remains an open challenge. Thresholds for output flagging and behavioral monitoring require empirical tuning and human oversight to balance false positives against security sensitivity. These issues are best addressed through iterative refinement rather than static rules. More broadly, we intentionally leave many operational details—such as review timelines, appeals processes, and specific criteria for institutional trust—to individual implementors. Different hosting entities face different     threat models, resource constraints, and user populations, making overly prescriptive specifications counterproductive. The framework provides architectural principles while preserving flexibility for context-specific adaptation.

Lastly, a malicious actor with genuine institutional affiliation could use biological AI tools for legitimate purposes while gradually probing system boundaries or accumulating knowledge applicable to harmful ends. Tier III monitoring provides some protection against this pattern, but detecting slow, careful misuse remains challenging. No access control system fully solves the insider threat problem.

\section{Conclusion}

Biological AI tools have reached a capabilities threshold that demands governance infrastructure, yet existing content-based restrictions remain fundamentally inadequate for biology. Inspired by the financial sector's AML system, this paper has proposed a three-tier KYC framework that addresses this mismatch by shifting focus from predicting output danger—a task current systems cannot reliably perform—to verifying user identity and maintaining accountability through institutional anchors.

The framework's core insight adapts a proven model from financial compliance: layered defenses combining entry-point verification, continuous monitoring, and behavioral pattern analysis. Tier I leverages research institutions as trust anchors, creating accountability through institutional vouching. Tier II applies available output screening tools while acknowledging their limitations. Tier III detects concerning patterns over time through longitudinal behavioral analysis. Each tier provides independent security value while compensating for the others' weaknesses.

Critically, this approach can be implemented immediately. Hosting entities can adopt institutional gatekeeping unilaterally through terms of service without waiting for regulatory mandates. The framework builds on existing institutional infrastructure rather. For legitimate researchers, friction should be minimal, and for potential bad actors, the framework raises costs through attribution, monitoring, and enforceable consequences.

Important limitations remain. Open-weight models bypass access control entirely; well-resourced state actors could establish front institutions; and the calibration of screening thresholds and behavioral patterns will require empirical tuning to balance false positives against security sensitivity. 

Despite these limitations, the framework offers substantial security gains over the status quo. The alternative is either restricting access so heavily that legitimate research suffers, or maintaining open access with minimal safeguards as capabilities grow more dangerous. This framework charts a middle path: preserving access for accountable researchers while raising barriers to misuse through institutional responsibility, continuous monitoring, and enforceable consequences.

As biological AI capabilities continue advancing, governance infrastructure will become increasingly critical. The framework proposed here provides a concrete starting point—implementable today, scalable as capabilities evolve, and grounded in the practical realities of both biological research and risk management. The question is not whether such infrastructure will eventually be necessary, but whether we implement it proactively or reactively. We advocate for the former.

\bibliographystyle{unsrtnat}  
\bibliography{Shakhnovich}

@inproceedings{10.1145/3375627.3375873,
	title        = {Monitoring Misuse for Accountable 'Artificial Intelligence as a Service'},
	author       = {Javadi, Seyyed Ahmad and Cloete, Richard and Cobbe, Jennifer and Lee, Michelle Seng Ah and Singh, Jatinder},
	year         = {2020},
	booktitle    = {Proceedings of the AAAI/ACM Conference on AI, Ethics, and Society},
	location     = {New York, NY, USA},
	publisher    = {Association for Computing Machinery},
	address      = {New York, NY, USA},
	series       = {AIES '20},
	pages        = {300–306},
	doi          = {10.1145/3375627.3375873},
	isbn         = {9781450371100},
	url          = {https://doi.org/10.1145/3375627.3375873},
	numpages     = {7}
}

@article{10.1242/dmm.052218,
	title        = {Making sense of missense: challenges and opportunities in variant pathogenicity prediction},
	author       = {Molotkov, Ivan and Mardis, Elaine R. and Artomov, Mykyta},
	year         = {2024},
	month        = {12},
	journal      = {Disease Models \& Mechanisms},
	volume       = {17},
	number       = {12},
	pages        = {dmm052218},
	doi          = {10.1242/dmm.052218},
	issn         = {1754-8403},
	url          = {https://doi.org/10.1242/dmm.052218},
	eprint       = {https://journals.biologists.com/dmm/article-pdf/17/12/dmm052218/3587958/dmm052218.pdf}
}

@misc{ackerman_biothreat_2025,
	title        = {Biothreat Benchmark Generation Framework for Evaluating Frontier AI Models I: The Task-Query Architecture},
	shorttitle   = {Biothreat {Benchmark} {Generation} {Framework} for {Evaluating} {Frontier} {AI} {Models} {I}},
	author       = {Ackerman, Gary and Behlendorf, Brandon and Kallenborn, Zachary and Almakki, Sheriff and Clifford, Doug and LaTourette, Jenna and Peterson, Hayley and Sheinbaum, Noah and Shoemaker, Olivia and Wetzel, Anna},
	year         = {2025},
	month        = dec,
	publisher    = {arXiv},
	doi          = {10.48550/arXiv.2512.08130},
	url          = {http://arxiv.org/abs/2512.08130},
	urldate      = {2026-01-17},
	note         = {arXiv:2512.08130 [cs]}
}

@article{af3complex,
	title        = {AF3Complex yields improved structural predictions of protein complexes},
	author       = {Feldman, Jonathan and Skolnick, Jeffrey},
	year         = {2025},
	month        = {07},
	journal      = {Bioinformatics},
	volume       = {41},
	number       = {8},
	pages        = {btaf432},
	doi          = {10.1093/bioinformatics/btaf432},
	issn         = {1367-4811},
	url          = {https://doi.org/10.1093/bioinformatics/btaf432},
	eprint       = {https://academic.oup.com/bioinformatics/article-pdf/41/8/btaf432/63901325/btaf432.pdf}
}

@misc{af3server_alphafold,
	title        = {AlphaFold Server},
	url          = {https://alphafoldserver.com/},
	urldate      = {2025-08-18}
}

@misc{americas_ai_action_plan_2025,
	title        = {America's AI Action Plan},
	author       = {The White House},
	year         = {2025},
	month        = jul,
	note         = {Section ``Invest in Biosecurity'' calls for mandatory nucleic acid synthesis screening, customer verification, data sharing among providers, and enforcement mechanisms tied to federal research funding; accessed 2026-01-08},
	howpublished = {\url{https://www.whitehouse.gov/wp-content/uploads/2025/07/Americas-AI-Action-Plan.pdf}}
}

@misc{anderljung2023protectingsocietyaimisuse,
	title        = {Protecting Society from AI Misuse: When are Restrictions on Capabilities Warranted?},
	author       = {Markus Anderljung and Julian Hazell},
	year         = {2023},
	url          = {https://arxiv.org/abs/2303.09377},
	eprint       = {2303.09377},
	archiveprefix = {arXiv},
	primaryclass = {cs.AI}
}

@misc{ball2025impossibilityseparatingintelligencejudgment,
	title        = {On the Impossibility of Separating Intelligence from Judgment: The Computational Intractability of Filtering for AI Alignment},
	author       = {Sarah Ball and Greg Gluch and Shafi Goldwasser and Frauke Kreuter and Omer Reingold and Guy N. Rothblum},
	year         = {2025},
	url          = {https://arxiv.org/abs/2507.07341},
	eprint       = {2507.07341},
	archiveprefix = {arXiv},
	primaryclass = {cs.AI}
}

@misc{bcbs_consolidated_kyc_2004,
	title        = {Consolidated KYC Risk Management},
	author       = {Basel Committee on Banking Supervision},
	year         = {2004},
	month        = oct,
	note         = {Bank for International Settlements. Accessed 2026-01-08},
	howpublished = {\url{https://www.bis.org/publ/bcbs110.pdf}}
}

@misc{black2025openweightgenomelanguagemodel,
	title        = {Open-weight genome language model safeguards: Assessing robustness via adversarial fine-tuning},
	author       = {James R. M. Black and Moritz S. Hanke and Aaron Maiwald and Tina Hernandez-Boussard and Oliver M. Crook and Jaspreet Pannu},
	year         = {2025},
	url          = {https://arxiv.org/abs/2511.19299},
	eprint       = {2511.19299},
	archiveprefix = {arXiv},
	primaryclass = {cs.LG}
}

@article{BRAVOURQUIZA2021101344,
	title        = {Does compliance with corporate governance codes help to mitigate financial distress?},
	author       = {Francisco Bravo-Urquiza and Elena Moreno-Ureba},
	year         = {2021},
	journal      = {Research in International Business and Finance},
	volume       = {55},
	pages        = {101344},
	doi          = {https://doi.org/10.1016/j.ribaf.2020.101344},
	issn         = {0275-5319},
	url          = {https://www.sciencedirect.com/science/article/pii/S0275531920309521}
}

@misc{brixi_genome_2025,
	title        = {Genome modeling and design across all domains of life with Evo 2},
	author       = {Brixi, Garyk and Durrant, Matthew G. and Ku, Jerome and Poli, Michael and Brockman, Greg and Chang, Daniel and Gonzalez, Gabriel A. and King, Samuel H. and Li, David B. and Merchant, Aditi T. and Naghipourfar, Mohsen and Nguyen, Eric and Ricci-Tam, Chiara and Romero, David W. and Sun, Gwanggyu and Taghibakshi, Ali and Vorontsov, Anton and Yang, Brandon and Deng, Myra and Gorton, Liv and Nguyen, Nam and Wang, Nicholas K. and Adams, Etowah and Baccus, Stephen A. and Dillmann, Steven and Ermon, Stefano and Guo, Daniel and Ilango, Rajesh and Janik, Ken and Lu, Amy X. and Mehta, Reshma and Mofrad, Mohammad R. K. and Ng, Madelena Y. and Pannu, Jaspreet and R\'{e}, Christopher and Schmok, Jonathan C. and John, John St and Sullivan, Jeremy and Zhu, Kevin and Zynda, Greg and Balsam, Daniel and Collison, Patrick and Costa, Anthony B. and Hernandez-Boussard, Tina and Ho, Eric and Liu, Ming-Yu and McGrath, Thomas and Powell, Kimberly and Burke, Dave P. and Goodarzi, Hani and Hsu, Patrick D. and Hie, Brian L.},
	year         = {2025},
	month        = feb,
	publisher    = {bioRxiv},
	doi          = {10.1101/2025.02.18.638918},
	url          = {https://www.biorxiv.org/content/10.1101/2025.02.18.638918v1},
	urldate      = {2025-08-17},
	copyright    = {\textcopyright{} 2025, Posted by Cold Spring Harbor Laboratory. This pre-print is available under a Creative Commons License (Attribution-NoDerivs 4.0 International), CC BY-ND 4.0, as described at http://creativecommons.org/licenses/by-nd/4.0/},
	note         = {Pages: 2025.02.18.638918 Section: New Results},
	language     = {en}
}

@misc{bu_export_control_manual,
	title        = {Export Control Manual},
	author       = {Boston University Office of Research Ethics \& Compliance},
	year         = {2023},
	url          = {https://www.bu.edu/research/ethics-compliance/research-security/export-control/export-control-manual/}
}

@article{cao_prediction_2025,
	title        = {Prediction of pathogenic mutations in human transmembrane proteins and their associated diseases via utilizing pre-trained Bio-LLMs},
	author       = {Cao, Lexin and Quan, Lijun and Chen, Qiufeng and Zhang, Bei and Zhang, Zhijun and Peng, Liangchen and Wang, Junkai and Jiang, Yelu and Nie, Liangpeng and Li, Geng and Wu, Tingfang and Lyu, Qiang},
	year         = {2025},
	month        = jul,
	journal      = {Communications Biology},
	volume       = {8},
	pages        = {1050},
	doi          = {10.1038/s42003-025-08452-7},
	issn         = {2399-3642},
	url          = {https://pmc.ncbi.nlm.nih.gov/articles/PMC12264167/},
	urldate      = {2025-10-29},
	pmid         = {40665056},
	pmcid        = {PMC12264167}
}

@article{chen_enhancing_nodate,
	title        = {Enhancing privacy in biosecurity with watermarked protein design},
	author       = {Chen, Yanshuo and Hu, Zhengmian and Wu, Yihan and Chen, Ruibo and Jin, Yongrui and Zhan, Marcus and Xie, Chengjin and Chen, Wei and Huang, Heng},
	url          = {https://dx.doi.org/10.1093/bioinformatics/btaf141},
	urldate      = {2026-01-17},
	language     = {en}
}

@article{chen_you_2020,
	title        = {Do you know your customer? Bank risk assessment based on machine learning},
	shorttitle   = {Do you know your customer?},
	author       = {Chen, Ting-Hsuan},
	year         = {2020},
	month        = jan,
	journal      = {Applied Soft Computing},
	volume       = {86},
	pages        = {105779},
	doi          = {10.1016/j.asoc.2019.105779},
	issn         = {1568-4946},
	url          = {https://www.sciencedirect.com/science/article/pii/S1568494619305605},
	urldate      = {2026-01-07}
}

@article{CHEN2023706,
	title        = {Learning protein fitness landscapes with deep mutational scanning data from multiple sources},
	author       = {Lin Chen and Zehong Zhang and Zhenghao Li and Rui Li and Ruifeng Huo and Lifan Chen and Dingyan Wang and Xiaomin Luo and Kaixian Chen and Cangsong Liao and Mingyue Zheng},
	year         = {2023},
	journal      = {Cell Systems},
	volume       = {14},
	number       = {8},
	pages        = {706--721.e5},
	doi          = {https://doi.org/10.1016/j.cels.2023.07.003},
	issn         = {2405-4712},
	url          = {https://www.sciencedirect.com/science/article/pii/S2405471223002107}
}

@misc{chukwudi_how_2025,
	title        = {How Enhanced Compliance Frameworks Reduce Litigation Risk and Improve Investor Confidence},
	author       = {Chukwudi, Cephas Ndubisi},
	year         = {2025},
	month        = oct,
	publisher    = {Social Science Research Network},
	address      = {Rochester, NY},
	doi          = {10.2139/ssrn.5700782},
	url          = {https://papers.ssrn.com/abstract=5700782},
	urldate      = {2026-01-08},
	type         = {{SSRN} {Scholarly} {Paper}},
	language     = {en}
}

@article{de_haro_biosecurity_2024,
	title        = {Biosecurity Risk Assessment for the Use of Artificial Intelligence in Synthetic Biology},
	author       = {De Haro, Leyma P.},
	year         = {2024},
	month        = jun,
	journal      = {Applied Biosafety: Journal of the American Biological Safety Association},
	volume       = {29},
	number       = {2},
	pages        = {96--107},
	doi          = {10.1089/apb.2023.0031},
	issn         = {1535-6760},
	url          = {https://pmc.ncbi.nlm.nih.gov/articles/PMC11313549/},
	urldate      = {2026-01-17},
	pmid         = {39131181},
	pmcid        = {PMC11313549}
}

@article{deepak_review_nodate,
	title        = {A Review on Know Your Customer (KYC) SystemUsing Blockchain Technology},
	author       = {Deepak, N and Patel, Joy and Jain, Likith and Srivastava, Mihika},
	language     = {en}
}

@article{dianzhuo_wang_without_2025,
	title        = {Without Safeguard, AI-Bio Integration Risks Accelerating Future Pandemics},
	author       = {Wang, Dianzhuo and Huot, Marian and Zhang, Zechen and Jiang, Kaiyi and Shakhnovich, Eugene and Esvelt, Kevin M},
	journal      = {Frontiers in Microbiology},
	url          = {https://www.frontiersin.org/journals/microbiology/articles/10.3389/fmicb.2025.1734561/abstract}
}

@misc{doc_entity_list,
	title        = {Entity List},
	author       = {U.S. Department of Commerce, Bureau of Industry and Security},
	note         = {Identifies foreign persons and entities subject to specific license requirements for exports, reexports, and transfers due to national security or foreign policy concerns; accessed 2026-01-08},
	howpublished = {\url{https://www.bis.gov/regulations/ear/744\#section-744.16}}
}

@article{doi:10.1126/science.ado1671,
	title        = {Protein design meets biosecurity},
	author       = {David Baker  and George Church},
	year         = {2024},
	journal      = {Science},
	volume       = {383},
	number       = {6681},
	pages        = {349--349},
	doi          = {10.1126/science.ado1671},
	url          = {https://www.science.org/doi/abs/10.1126/science.ado1671},
	eprint       = {https://www.science.org/doi/pdf/10.1126/science.ado1671}
}

@article{doi:10.1126/science.adu8578,
	title        = {Strengthening nucleic acid biosecurity screening against generative protein design tools},
	author       = {Bruce J. Wittmann  and Tessa Alexanian  and Craig Bartling  and Jacob Beal  and Adam Clore  and James Diggans  and Kevin Flyangolts  and Bryan T. Gemler  and Tom Mitchell  and Steven T. Murphy  and Nicole E. Wheeler  and Eric Horvitz},
	year         = {2025},
	journal      = {Science},
	volume       = {390},
	number       = {6768},
	pages        = {82--87},
	doi          = {10.1126/science.adu8578},
	url          = {https://www.science.org/doi/abs/10.1126/science.adu8578},
	eprint       = {https://www.science.org/doi/pdf/10.1126/science.adu8578}
}

@misc{ecfr_42cfr73_select_agents_2025,
	title        = {42 CFR Part 73 -- Select Agents and Toxins},
	author       = {U.S. Government Publishing Office},
	year         = {2025},
	url          = {https://www.ecfr.gov/current/title-42/chapter-I/subchapter-F/part-73},
	howpublished = {Electronic Code of Federal Regulations}
}

@misc{egan_oversight_2023,
	title        = {Oversight for Frontier AI through a Know-Your-Customer Scheme for Compute Providers},
	author       = {Egan, Janet and Heim, Lennart},
	year         = {2023},
	month        = oct,
	publisher    = {arXiv},
	doi          = {10.48550/arXiv.2310.13625},
	url          = {http://arxiv.org/abs/2310.13625},
	urldate      = {2026-01-07},
	note         = {arXiv:2310.13625 [cs]},
	language     = {en}
}

@misc{esm_faqs,
	title        = {Evolutionary Scale FAQs},
	url          = {https://forge.evolutionaryscale.ai/faq},
	urldate      = {2025-08-18}
}

@misc{facini_biosecurity_2025,
	title        = {Biosecurity in the Next Administration - The Council on Strategic Risks},
	author       = {Facini, Andrew},
	year         = {2025},
	month        = jan,
	url          = {https://councilonstrategicrisks.org/2025/01/31/biosecurity-in-the-next-administration/},
	urldate      = {2026-01-07},
	language     = {en-US}
}

@misc{fatf_recommendations_2025,
	title        = {International Standards on Combating Money Laundering and the Financing of Terrorism \& Proliferation: The FATF Recommendations},
	author       = {Financial Action Task Force (FATF)},
	year         = {2025},
	month        = jun,
	note         = {Updated June 2025. Describes the risk-based approach, including enhanced due diligence for higher-risk customers; accessed 2026-01-08},
	howpublished = {\url{https://www.fatf-gafi.org/en/publications/Fatfrecommendations/Fatf-recommendations.html}}
}

@misc{fatf_risk_based_approach,
	title        = {Guidance on the Risk-Based Approach: The Banking Sector},
	author       = {Financial Action Task Force (FATF)},
	note         = {Explains that AML/CFT controls should be proportionate to risk and avoid unnecessary disruption of legitimate activity; accessed 2026-01-08},
	howpublished = {\url{https://www.fatf-gafi.org/content/dam/fatf-gafi/guidance/Risk-Based-Approach-Banking-Sector.pdf}}
}

@misc{fed_payments_study_frps,
	title        = {Federal Reserve Payments Study (FRPS)},
	author       = {Federal Reserve Board},
	note         = {Official U.S. Federal Reserve Payments Study tracking aggregate noncash payment volumes in the U.S.; accessed 2026-01-08},
	howpublished = {\url{https://www.federalreserve.gov/paymentsystems/fr-payments-study.htm}}
}

@misc{fed_payments_study_npips_2021_22,
	title        = {National Payment Volumes, Detailed Data – NPIPS (CY 2021 and 2022)},
	author       = {Federal Reserve Board},
	note         = {Detailed noncash payment volume data from the Federal Reserve Payments Study; accessed 2026-01-08},
	howpublished = {\url{https://www.federalreserve.gov/paymentsystems/2024-November-The-Federal-Reserve-Payments-Study.htm}}
}

@misc{fed_supervision_manual_kyc,
	title        = {Bank Secrecy Act: ``Know Your Customer''},
	author       = {Board of Governors of the Federal Reserve System},
	note         = {Federal Reserve Supervision Manual. Accessed 2026-01-08},
	howpublished = {\url{https://www.federalreserve.gov/boarddocs/SupManual/bsa/bsa\_p5.pdf}}
}

@article{feldman_opinion_2025,
	title        = {Opinion {\textbar} AI just created a working virus. The U.S. isn't prepared for that.},
	author       = {Feldman, Tal and Feldman, Jonathan},
	year         = {2025},
	month        = sep,
	journal      = {The Washington Post},
	issn         = {0190-8286},
	url          = {https://www.washingtonpost.com/opinions/2025/09/25/artificial-intelligence-advance-virus-created/},
	urldate      = {2025-10-21},
	language     = {en-US}
}

@article{feldman_position_nodate,
	title        = {Position: Restricted Release of Advanced Biological Models Safeguards Biosecurity},
	author       = {Feldman, Jonathan and Feldman, Tal},
	url          = {https://openreview.net/forum?id=9Nhrk0Cuk0},
	language     = {en}
}

@misc{feldman_resilient_2025,
	title        = {Resilient Biosecurity in the Era of AI-Enabled Bioweapons},
	author       = {Feldman, Jonathan and Feldman, Tal},
	year         = {2025},
	month        = aug,
	publisher    = {arXiv},
	doi          = {10.48550/arXiv.2509.02610},
	url          = {http://arxiv.org/abs/2509.02610},
	urldate      = {2025-10-21},
	note         = {arXiv:2509.02610 [q-bio]},
	eprint       = {2509.02610},
	archiveprefix = {arXiv},
	primaryclass = {q-bio.QM}
}

@misc{ffiec_authentication_access_2021,
	title        = {Authentication and Access to Financial Institution Services and Systems},
	author       = {Federal Financial Institutions Examination Council (FFIEC)},
	year         = {2021},
	month        = aug,
	note         = {Guidance emphasizing monitoring, logging, and reporting of activities to identify and track unauthorized access; accessed 2026-01-08},
	howpublished = {\url{https://www.ffiec.gov/sites/default/files/media/press-releases/2021/authentication-and-access-to-financial-institution-services-and-systems.pdf}}
}

@misc{ffiec_bsaaml_manual,
	title        = {BSA/AML Examination Manual},
	author       = {Federal Financial Institutions Examination Council (FFIEC)},
	note         = {Accessed 2026-01-08},
	howpublished = {\url{https://bsaaml.ffiec.gov/manual}}
}

@misc{ffiec_sar_overview,
	title        = {BSA/AML Examination Manual: Suspicious Activity Reporting--Overview},
	author       = {Federal Financial Institutions Examination Council (FFIEC)},
	note         = {Explains SARs as a core regulatory reporting mechanism for sharing information about suspicious actors and activity; accessed 2026-01-08},
	howpublished = {\url{https://bsaaml.ffiec.gov/manual/AssessingComplianceWithBSARegulatoryRequirements/04}}
}

@misc{fincen_314b,
	title        = {Section 314(b) Information Sharing},
	author       = {Financial Crimes Enforcement Network (FinCEN)},
	note         = {Describes voluntary information sharing among financial institutions regarding suspected money laundering and terrorist financing; accessed 2026-01-08},
	howpublished = {\url{https://www.fincen.gov/resources/section-314b}}
}

@misc{fincen_cdd_final_rule_fedreg_2016,
	title        = {Customer Due Diligence Requirements for Financial Institutions},
	author       = {Financial Crimes Enforcement Network (FinCEN)},
	year         = {2016},
	note         = {31 CFR Parts 1010, 1020, 1023, 1024, and 1026},
	howpublished = {\url{https://www.federalregister.gov/documents/2016/05/11/2016-10567/customer-due-diligence-requirements-for-financial-institutions}}
}

@misc{finra_rule_2090_kyc,
	title        = {FINRA Rule 2090: Know Your Customer},
	author       = {Financial Industry Regulatory Authority (FINRA)},
	note         = {Accessed 2026-01-08},
	howpublished = {\url{https://www.finra.org/rules-guidance/rulebooks/finra-rules/2090}}
}

@report{fsap_annual_report_2024,
	title        = {2024 Annual Report of the Federal Select Agent Program},
	year         = {2024},
	url          = {https://www.selectagents.gov/resources/publications/docs/2024-FSAP-Annual-Report\%5F508.pdf},
	institution  = {Centers for Disease Control and Prevention and U.S. Department of Agriculture}
}

@report{fsap_inspection_summary_2024,
	title        = {2024 FSAP Inspection Report Processing Annual Summary},
	year         = {2024},
	url          = {https://www.selectagents.gov/resources/publications/docs/2024-FSAP-Inspection-Report-Processing-Annual-Summary\%5F508.pdf},
	institution  = {Centers for Disease Control and Prevention and U.S. Department of Agriculture}
}

@misc{fsap_security_risk_assessment_2024,
	title        = {Security Risk Assessments},
	author       = {Federal Select Agent Program},
	year         = {2024},
	url          = {https://www.selectagents.gov/compliance/risk.htm},
	note         = {Describes how entities submit individuals for federal security risk assessments and are responsible for authorizing and controlling access based on results.},
	howpublished = {Centers for Disease Control and Prevention and USDA}
}

@article{gao_biosafety_2025,
	title        = {Biosafety concept: Origins, Evolution, and Prospects},
	shorttitle   = {Biosafety concept},
	author       = {Gao, Wanying and Wu, Zongzhen and Zuo, Kunlan and Xiang, Qiangyu and Chen, Xiaoya and Zhang, Lu and Liu, Huan},
	year         = {2025},
	month        = jul,
	journal      = {Biosafety and Health},
	volume       = {7},
	number       = {4},
	pages        = {209--217},
	doi          = {10.1016/j.bsheal.2025.07.003},
	issn         = {2096-6962},
	url          = {https://pmc.ncbi.nlm.nih.gov/articles/PMC12412398/},
	urldate      = {2026-01-07},
	pmid         = {40918204},
	pmcid        = {PMC12412398}
}

@article{grady_institutional_2015,
	title        = {Institutional Review Boards},
	author       = {Grady, Christine},
	year         = {2015},
	month        = nov,
	journal      = {Chest},
	volume       = {148},
	number       = {5},
	pages        = {1148--1155},
	doi          = {10.1378/chest.15-0706},
	issn         = {0012-3692},
	url          = {https://pmc.ncbi.nlm.nih.gov/articles/PMC4631034/},
	urldate      = {2026-01-07},
	pmid         = {26042632},
	pmcid        = {PMC4631034}
}

@misc{hayes_simulating_2024,
	title        = {Simulating 500 million years of evolution with a language model},
	author       = {Hayes, Thomas and Rao, Roshan and Akin, Halil and Sofroniew, Nicholas J. and Oktay, Deniz and Lin, Zeming and Verkuil, Robert and Tran, Vincent Q. and Deaton, Jonathan and Wiggert, Marius and Badkundri, Rohil and Shafkat, Irhum and Gong, Jun and Derry, Alexander and Molina, Raul S. and Thomas, Neil and Khan, Yousuf and Mishra, Chetan and Kim, Carolyn and Bartie, Liam J. and Nemeth, Matthew and Hsu, Patrick D. and Sercu, Tom and Candido, Salvatore and Rives, Alexander},
	year         = {2024},
	month        = jul,
	publisher    = {bioRxiv},
	doi          = {10.1101/2024.07.01.600583},
	url          = {https://www.biorxiv.org/content/10.1101/2024.07.01.600583v1},
	urldate      = {2025-06-04},
	copyright    = {\textcopyright{} 2024, Posted by Cold Spring Harbor Laboratory. This pre-print is available under a Creative Commons License (Attribution-NonCommercial-NoDerivs 4.0 International), CC BY-NC-ND 4.0, as described at http://creativecommons.org/licenses/by-nc-nd/4.0/},
	note         = {Pages: 2024.07.01.600583 Section: New Results},
	language     = {en}
}

@misc{hhs_synthetic_nucleic_acid_guidance_2023,
	title        = {Screening Framework Guidance for Providers and Users of Synthetic Nucleic Acids},
	author       = {Department of Health and Human Services, Administration for Strategic Preparedness and Response},
	year         = {2023},
	note         = {Provides baseline standards for screening synthetic nucleic acid orders and verifying customer identity to address biosecurity concerns; accessed 2026-01-08},
	howpublished = {\url{https://www.federalregister.gov/documents/2023/10/13/2023-22540/screening-framework-guidance-for-providers-and-users-of-synthetic-nucleic-acids}}
}

@article{hie_evolutionary_2022,
	title        = {Evolutionary velocity with protein language models predicts evolutionary dynamics of diverse proteins},
	author       = {Hie, Brian L. and Yang, Kevin K. and Kim, Peter S.},
	year         = {2022},
	month        = apr,
	journal      = {Cell Systems},
	volume       = {13},
	number       = {4},
	pages        = {274--285.e6},
	doi          = {10.1016/j.cels.2022.01.003},
	issn         = {2405-4720},
	language     = {eng},
	pmid         = {35120643}
}

@article{HOFFMANN2023106165,
	title        = {Safety by design: Biosafety and biosecurity in the age of synthetic genomics},
	author       = {Stefan A. Hoffmann and James Diggans and Douglas Densmore and Junbiao Dai and Tom Knight and Emily Leproust and Jef D. Boeke and Nicole Wheeler and Yizhi Cai},
	year         = {2023},
	journal      = {iScience},
	volume       = {26},
	number       = {3},
	pages        = {106165},
	doi          = {https://doi.org/10.1016/j.isci.2023.106165},
	issn         = {2589-0042},
	url          = {https://www.sciencedirect.com/science/article/pii/S2589004223002420}
}

@article{Huot2025.05.12.653592,
	title        = {Generative model of SARS-CoV-2 variants under functional and immune pressure unveils viral escape potential and antibody resilience},
	author       = {Huot, Marian and Rosenbaum, Pierre and Planchais, Cyril and Mouquet, Hugo and Monasson, R{\'e}mi and Cocco, Simona},
	year         = {2025},
	journal      = {bioRxiv},
	publisher    = {Cold Spring Harbor Laboratory},
	pages        = {2025--05},
	doi          = {10.1101/2025.05.12.653592},
	url          = {https://www.biorxiv.org/content/early/2025/05/13/2025.05.12.653592},
	elocation-id = {2025.05.12.653592},
	eprint       = {https://www.biorxiv.org/content/early/2025/05/13/2025.05.12.653592.full.pdf}
}

@article{huss_mapping_2021,
	title        = {Mapping the functional landscape of the receptor binding domain of T7 bacteriophage by deep mutational scanning},
	author       = {Huss, Phil and Meger, Anthony and Leander, Megan and Nishikawa, Kyle and Raman, Srivatsan},
	year         = {2021},
	month        = mar,
	journal      = {eLife},
	volume       = {10},
	pages        = {e63775},
	doi          = {10.7554/eLife.63775},
	issn         = {2050-084X},
	url          = {https://doi.org/10.7554/eLife.63775},
	urldate      = {2025-06-26},
	note         = {Publisher: eLife Sciences Publications, Ltd},
	editor       = {Storz, Gisela and Turnbaugh, Peter and Debarbieux, Laurent and Duerkop, Breck A and Fraser, James S}
}

@article{hwang_genomic_2024,
	title        = {Genomic language model predicts protein co-regulation and function},
	author       = {Hwang, Yunha and Cornman, Andre L. and Kellogg, Elizabeth H. and Ovchinnikov, Sergey and Girguis, Peter R.},
	year         = {2024},
	month        = apr,
	journal      = {Nature Communications},
	volume       = {15},
	number       = {1},
	pages        = {2880},
	doi          = {10.1038/s41467-024-46947-9},
	issn         = {2041-1723},
	url          = {https://www.nature.com/articles/s41467-024-46947-9},
	urldate      = {2025-08-05},
	copyright    = {2024 The Author(s)},
	note         = {Publisher: Nature Publishing Group},
	language     = {en}
}

@misc{ibm_aml_transaction_monitoring,
	title        = {What Is AML Transaction Monitoring?},
	author       = {IBM},
	note         = {Describes continuous monitoring of customer transactions to detect anomalous or inconsistent behavior; accessed 2026-01-08},
	howpublished = {\url{https://www.ibm.com/think/topics/aml-transaction-monitoring}}
}

@article{King2025.09.12.675911,
	title        = {Generative design of novel bacteriophages with genome language models},
	author       = {King, Samuel H. and Driscoll, Claudia L. and Li, David B. and Guo, Daniel and Merchant, Aditi T. and Brixi, Garyk and Wilkinson, Max E. and Hie, Brian L.},
	year         = {2025},
	journal      = {bioRxiv},
	publisher    = {Cold Spring Harbor Laboratory},
	doi          = {10.1101/2025.09.12.675911},
	url          = {https://www.biorxiv.org/content/early/2025/09/17/2025.09.12.675911},
	elocation-id = {2025.09.12.675911},
	eprint       = {https://www.biorxiv.org/content/early/2025/09/17/2025.09.12.675911.full.pdf}
}

@article{LIU2025100964,
	title        = {Discovering topics and trends in biosecurity law research: A machine learning approach},
	author       = {Yang Liu},
	year         = {2025},
	journal      = {One Health},
	volume       = {20},
	pages        = {100964},
	doi          = {https://doi.org/10.1016/j.onehlt.2024.100964},
	issn         = {2352-7714},
	url          = {https://www.sciencedirect.com/science/article/pii/S2352771424002908}
}

@article{marsoof_content-filtering_2023,
	title        = {Content-filtering AI systems–limitations, challenges and regulatory approaches},
	author       = {Marsoof, Althaf and Luco, Andr\'{e}s and Tan, Harry and Joty, Shafiq},
	year         = {2023},
	month        = jan,
	journal      = {Information \& Communications Technology Law},
	volume       = {32},
	number       = {1},
	pages        = {64--101},
	doi          = {10.1080/13600834.2022.2078395},
	issn         = {1360-0834},
	url          = {https://www.tandfonline.com/doi/citedby/10.1080/13600834.2022.2078395},
	urldate      = {2026-01-08},
	note         = {Publisher: Routledge}
}

@article{miller_post-processing_1997,
	title        = {Post-processing of BLAST results using databases of clustered sequences},
	author       = {Miller, G. S. and Fuchs, R.},
	year         = {1997},
	month        = feb,
	journal      = {Bioinformatics},
	volume       = {13},
	number       = {1},
	pages        = {81--87},
	doi          = {10.1093/bioinformatics/13.1.81},
	issn         = {1367-4803},
	url          = {https://doi.org/10.1093/bioinformatics/13.1.81},
	urldate      = {2025-08-11}
}

@misc{mit_restricted_party,
	title        = {Restricted Party Screening},
	author       = {Massachusetts Institute of Technology Office of the Vice President for Research},
	year         = {2025},
	url          = {https://research.mit.edu/integrity-and-compliance/export-control/key-concepts/restricted-party-screening}
}

@misc{moulange2023responsiblegovernancebiologicaldesign,
	title        = {Towards Responsible Governance of Biological Design Tools},
	author       = {Richard Moulange and Max Langenkamp and Tessa Alexanian and Samuel Curtis and Morgan Livingston},
	year         = {2023},
	url          = {https://arxiv.org/abs/2311.15936},
	eprint       = {2311.15936},
	archiveprefix = {arXiv},
	primaryclass = {cs.CY}
}

@incollection{MOUNEYRAC2017241,
	title        = {Chapter 9 - The Role of Laboratory Experiments in the Validation of Field Data},
	author       = {Catherine Mouneyrac and Fabienne Lagarde and Am\'{e}lie Ch\^{a}tel and Farhan R. Khan and Kristian Syberg and Annemette Palmqvist},
	year         = {2017},
	booktitle    = {Characterization and Analysis of Microplastics},
	publisher    = {Elsevier},
	series       = {Comprehensive Analytical Chemistry},
	volume       = {75},
	pages        = {241--273},
	doi          = {https://doi.org/10.1016/bs.coac.2016.10.005},
	issn         = {0166-526X},
	url          = {https://www.sciencedirect.com/science/article/pii/S0166526X16301544},
	editor       = {Teresa A.P. Rocha-Santos and Armando C. Duarte}
}

@article{NEURIPS2021_f51338d7,
	title        = {Language models enable zero-shot prediction of the effects of mutations on protein function},
	author       = {Meier, Joshua and Rao, Roshan and Verkuil, Robert and Liu, Jason and Sercu, Tom and Rives, Alex},
	year         = {2021},
	booktitle    = {Advances in Neural Information Processing Systems},
	publisher    = {Curran Associates, Inc.},
	volume       = {34},
	pages        = {29287--29303},
	url          = {https://proceedings.neurips.cc/paper\%5Ffiles/paper/2021/file/f51338d736f95dd42427296047067694-Paper.pdf},
	editor       = {M. Ranzato and A. Beygelzimer and Y. Dauphin and P.S. Liang and J. Wortman Vaughan}
}

@misc{nih_durc_policy,
	title        = {Dual Use Research of Concern (DURC) Policy},
	author       = {National Institutes of Health},
	note         = {Describes institutional oversight responsibilities for NIH-funded research involving dual-use concerns; accessed 2026-01-08},
	howpublished = {\url{https://aspr.hhs.gov/S3/Pages/Dual-Use-Research-of-Concern-Oversight-Policy-Framework.aspx/}}
}

@misc{nist_sp800_63b_4,
	title        = {SP 800-63B-4: Digital Identity Guidelines -- Authentication and Lifecycle Management},
	author       = {National Institute of Standards and Technology},
	year         = {2025},
	note         = {Includes risk indicators such as authentication from an unexpected geolocation or IP address block; accessed 2026-01-08},
	howpublished = {\url{https://csrc.nist.gov/pubs/sp/800/63/b/4/final}}
}

@misc{noauthor_fact_2022,
	title        = {FACT SHEET: The U.S. Government Funds Biotechnology and Biomanufacturing Innovation: Department of Energy},
	shorttitle   = {{FACT} {SHEET}},
	year         = {2022},
	month        = dec,
	journal      = {Energy.gov},
	url          = {https://www.energy.gov/eere/bioenergy/articles/fact-sheet-us-government-funds-biotechnology-and-biomanufacturing},
	urldate      = {2026-01-08},
	language     = {en}
}

@misc{noauthor_federal_nodate,
	title        = {Federal Funding of Research and Development {\textbar} Research Starters {\textbar} EBSCO Research},
	journal      = {EBSCO},
	url          = {https://www.ebsco.com},
	urldate      = {2026-01-08},
	language     = {en}
}

@article{noauthor_frameworks_2017,
	title        = {Frameworks},
	year         = {2017},
	month        = feb,
	journal      = {NIST},
	url          = {https://www.nist.gov/frameworks},
	urldate      = {2026-01-08},
	note         = {Last Modified: 2024-09-25T07:56-04:00},
	language     = {en}
}

@misc{noauthor_global_nodate,
	title        = {Global Risk Index for AI-enabled Biological Tools},
	url          = {https://www.rand.org/randeurope/research/projects/2024/ai-risk-index.html},
	urldate      = {2026-01-17},
	language     = {en}
}

@misc{noauthor_how_nodate,
	title        = {How can data-sharing help to spot scammers and prevent online fraud?},
	journal      = {Lloyds Banking Group},
	url          = {https://www.lloydsbankinggroup.com/insights/how-can-data-sharing-help-to-spot-scammers-and-prevent-online-fr.html},
	urldate      = {2026-01-08},
	language     = {en}
}

@misc{noauthor_institutional_2025,
	title        = {Institutional Biosafety Review},
	year         = {2025},
	month        = may,
	url          = {https://www.ibc.pitt.edu/institutional-biosafety-review},
	urldate      = {2026-01-07},
	language     = {en}
}

@article{noauthor_standards_2016,
	title        = {Standards},
	year         = {2016},
	month        = jun,
	journal      = {NIST},
	url          = {https://www.nist.gov/standards},
	urldate      = {2026-01-08},
	note         = {Last Modified: 2025-12-12T14:51-05:00},
	language     = {en}
}

@article{noauthor_user_nodate,
	title        = {User Awareness of Security Countermeasures and Its Impact on Information Systems Misuse: A Deterrence Approach},
	author       = {D'Arcy, John and Hovav, Anat and Galletta, Dennis},
	year         = {2009},
	journal      = {Information Systems Research},
	volume       = {20},
	number       = {1},
	pages        = {79--98},
	url          = {https://pubsonline.informs.org/doi/10.1287/isre.1070.0160},
	doi          = {10.1287/isre.1070.0160}
}

@misc{ofac_sdn_list,
	title        = {Specially Designated Nationals (SDN) List},
	author       = {U.S. Department of the Treasury, Office of Foreign Assets Control},
	note         = {Lists individuals and entities that U.S. persons are generally prohibited from dealing with under sanctions; accessed 2026-01-08},
	howpublished = {\url{https://www.treasury.gov/resource-center/sanctions/SDN-List/Pages/default.aspx}}
}

@misc{rochester_export_compliance,
	title        = {US Export Compliance},
	author       = {University of Rochester Office of Research Compliance},
	year         = {2025},
	url          = {https://www.rochester.edu/university-research/compliance/research-security/us-export-compliance/}
}

@article{ruffolo_designing_2024,
	title        = {Designing proteins with language models},
	author       = {Ruffolo, Jeffrey A. and Madani, Ali},
	year         = {2024},
	month        = feb,
	journal      = {Nature Biotechnology},
	volume       = {42},
	number       = {2},
	pages        = {200--202},
	doi          = {10.1038/s41587-024-02123-4},
	issn         = {1546-1696},
	url          = {https://doi.org/10.1038/s41587-024-02123-4}
}

@regulation{select_agent_regulations,
	title        = {Select Agent and Toxin Regulations},
	year         = {2024},
	note         = {42 C.F.R. Part 73; 7 C.F.R. Part 331; 9 C.F.R. Part 121},
	institution  = {U.S. Department of Health and Human Services; U.S. Department of Agriculture}
}

@article{serpico_institutional_2022,
	title        = {Institutional Review Board Use of Outside Experts: A National Survey},
	shorttitle   = {Institutional {Review} {Board} {Use} of {Outside} {Experts}},
	author       = {Serpico, Kimberley and Rahimzadeh, Vasiliki and Gelinas, Luke and Hartsmith, Lauren and Lynch, Holly Fernandez and Anderson, Emily E.},
	year         = {2022},
	journal      = {AJOB empirical bioethics},
	volume       = {13},
	number       = {4},
	pages        = {251--262},
	doi          = {10.1080/23294515.2022.2090459},
	issn         = {2329-4515},
	url          = {https://pmc.ncbi.nlm.nih.gov/articles/PMC10360021/},
	urldate      = {2026-01-07},
	pmid         = {35748820},
	pmcid        = {PMC10360021}
}

@article{Stark2025.11.20.689494,
	title        = {BoltzGen: Toward Universal Binder Design},
	author       = {Stark, Hannes and Faltings, Felix and Choi, MinGyu and Xie, Yuxin and Hur, Eunsu and O{\textquoteright}Donnell, Timothy and Bushuiev, Anton and U{\c c}ar, Talip and Passaro, Saro and Mao, Weian and Reveiz, Mateo and Bushuiev, Roman and Pluskal, Tom{\'a}{\v s} and Sivic, Josef and Kreis, Karsten and Vahdat, Arash and Ray, Shamayeeta and Goldstein, Jonathan T. and Savinov, Andrew and Hambalek, Jacob A. and Gupta, Anshika and Taquiri-Diaz, Diego A. and Zhang, Yaotian and Hatstat, A. Katherine and Arada, Angelika and Kim, Nam Hyeong and Tackie-Yarboi, Ethel and Boselli, Dylan and Schnaider, Lee and Liu, Chang C. and Li, Gene-Wei and Hnisz, Denes and Sabatini, David M. and DeGrado, William F. and Wohlwend, Jeremy and Corso, Gabriele and Barzilay, Regina and Jaakkola, Tommi},
	year         = {2025},
	journal      = {bioRxiv},
	publisher    = {Cold Spring Harbor Laboratory},
	doi          = {10.1101/2025.11.20.689494},
	url          = {https://www.biorxiv.org/content/early/2025/11/24/2025.11.20.689494},
	elocation-id = {2025.11.20.689494},
	eprint       = {https://www.biorxiv.org/content/early/2025/11/24/2025.11.20.689494.full.pdf}
}

@misc{state_dept_fto_list,
	title        = {U.S. Department of State Foreign Terrorist Organizations (FTO) List},
	author       = {U.S. Department of State},
	note         = {Maintains a list of designated terrorist organizations under U.S. law; accessed 2026-01-08},
	howpublished = {\url{https://www.state.gov/foreign-terrorist-organizations/}}
}

@article{suveena_translational_2025,
	title        = {The translational impact of bioinformatics on traditional wet lab techniques},
	author       = {Suveena, S. and Rekha, Akhiya Anilkumar and Rani, J. R. and V Oommen, Oommen and Ramakrishnan, Reshmi},
	year         = {2025},
	journal      = {Advances in Pharmacology (San Diego, Calif.)},
	volume       = {103},
	pages        = {287--311},
	doi          = {10.1016/bs.apha.2025.01.012},
	issn         = {1557-8925},
	language     = {eng},
	pmid         = {40175046}
}

@misc{turn0search2,
	title        = {What Is the Know Your Customer (KYC) Law?},
	note         = {Explains KYC identity verification requirements; accessed 2026-01-08},
	howpublished = {\url{https://www.sofi.com/learn/content/know-your-customer-kyc/}}
}

\end{document}